\documentclass[fleqn,usenatbib]{mnras}
\usepackage{longtable}
\usepackage{newtxtext,newtxmath}
\usepackage[T1]{fontenc}
\usepackage{ae,aecompl}
\usepackage{graphicx}	
\defcitealias{sfd}{SFD98}
\defcitealias{ccm89}{CCM89}
\defcitealias{schlafly2016}{SMS16}
\defcitealias{davenport2014}{DIB14}
\defcitealias{gould}{G09}
\defcitealias{av}{G12}
\defcitealias{g17}{G17}
\defcitealias{lallement2014}{LVV14}
\defcitealias{lallement2019}{LVV19}
\defcitealias{green2019}{GSF19}
\defcitealias{arenou}{AGG92}
\defcitealias{drimmel}{DCL03}
\defcitealias{wang2019}{WC19}
\defcitealias{2015ApJ...798...88M}{MF15}
\defcitealias{gm2020}{Paper I}

\title[{\it Gaia} DR2 giants in the Galactic dust]
{{\it Gaia} DR2 giants in the Galactic dust -- II. Application of the reddening maps and models.}

\author[G. A. Gontcharov and A. V. Mosenkov]{
George A. Gontcharov\thanks{E-mail: georgegontcharov@yahoo.com}
and Aleksandr V. Mosenkov
\\
Central Astronomical Observatory, Russian Academy of Sciences, 65/1 Pulkovskoye chaussee, St. Petersburg, 196140 Russia
}

\date{Accepted 2020 August 31; Received 2020 August 23; in original form 2020 April 9}

\pubyear{2020}

\begin{document}
\label{firstpage}
\pagerange{\pageref{firstpage}--\pageref{lastpage}}
\maketitle

\begin{abstract}
We exploit a complete sample of 101\,810 {\it Gaia} DR2 giants, selected in Paper I in the space cylinder with a radius of 700 pc around the Sun and a height of 
$|Z|=1800$ pc, using the {\it Gaia} DR2 parallaxes, $G_\mathrm{BP}$ and $G_\mathrm{RP}$ photometry, and {\it WISE} $W3$ photometry. We explain the spatial variations 
of the modes of the observables $G_\mathrm{BP}-G_\mathrm{RP}$ and $G_\mathrm{RP}-W3$ by the spatial variations of the corresponding reddenings described in the GM20 
3D dust distribution model. Presented in this paper, GM20 is an advanced version of the model introduced by Gontcharov in 2009. 
GM20 proposes two intersecting dust layers, along the Galactic mid-plane and in the Gould Belt, with exponential vertical and sinusoidal longitudinal variations of 
the dust spatial density in each layer. The Belt layer is an ellipse, oriented nearly between the centre and anticentre of the Galaxy, and with a semi-major and 
semi-minor axes of 600 and 146~pc, respectively. $G_\mathrm{BP}-G_\mathrm{RP}$ and $G_\mathrm{RP}-W3$ give similar solutions, but different equatorial layer scale 
heights of $150\pm15$ and $180\pm15$~pc, respectively, and $(G_\mathrm{BP}-G_\mathrm{RP})_0=(1.14\pm0.01)-(0.022\pm0.010)\,|Z|$, 
$(G_\mathrm{RP}-W3)_0=(1.44\pm0.01)-(0.015\pm0.010)\,|Z|$, where $Z$ is in kpc.
We compare GM20 with several 3D reddening models and maps in their ability to predict the observed colour modes.
GM20 and the 3D map by Gontcharov appear to be the best among the models and maps, respectively.
However, the most reliable models and maps mainly disagree only in their estimates of low reddening, including the reddening across the whole dust layer.
\end{abstract}


\section{Introduction}
\label{intro}

In our previous paper \citep[][hereafter Paper I]{gm2020} we verified the usage of the parallaxes and photometry from {\it Gaia} 
DR2 \citep{gaiabrown, gaiaevans} in order to simultaneously derive some key properties of the Galactic dust layer and clump 
giants embedded into or seen through this layer.

In \citetalias{gm2020} we selected a complete sample of 101\,810 {\it Gaia} DR2 giants in the giant clump domain of the 
Hertzsprung--Russell (HR) diagram within the space cylinder with a radius of 700 pc around the Sun and a height of $|Z|=1800$ pc 
along the $Z$ Galactic rectangular coordinate.
To select the sample, we used the {\it Gaia} DR2 parallaxes in combination with the photometry from the {\it Gaia} DR2 
$G_\mathrm{BP}$, $G_\mathrm{RP}$ bands and the $W3$ band from {\it Wide-field Infrared Survey Explorer (WISE}, allWISE, 
\citealt{wise}).

\citet{g2017} had shown that the modes of the intrinsic (dereddened) colours and absolute magnitudes of such a sample are 
completely defined by the clump giants\footnote{To obtain the modes, we follow \citet{g2017}, who rounds the observables up 
to 0.01 mag and finds the tops of their histograms in each spatial cell. In the case of multimodal histogram, the lowest value 
is selected, since it is more probable for the unreddened or slightly reddened clump.}.
The intrinsic colours and absolute magnitudes of the clump vary only with age and metallicity and can be presented in the space 
under consideration as some simple functions of $|Z|$ \citep{girardi2016, onaltas2016}.
Therefore, we assumed for our sample that the spatial variations of the modes of the observables
$G_\mathrm{BP}-G_\mathrm{RP}$, $G_\mathrm{RP}-W3$, $M_\mathrm{G_\mathrm{BP}}+A_\mathrm{G_\mathrm{BP}}$,
$M_\mathrm{G_\mathrm{RP}}+A_\mathrm{G_\mathrm{RP}}$, and $M_\mathrm{W3}+A_\mathrm{W3}$
reflect the spatial variations of the reddenings $E(G_\mathrm{BP}-G_\mathrm{RP})$ and $E(G_\mathrm{RP}-W3)$, 
extinctions $A_\mathrm{G_\mathrm{BP}}$, $A_\mathrm{G_\mathrm{RP}}$, and $A_\mathrm{W3}$
in a combination with some linear gradients of the dereddened colours $(G_\mathrm{BP}-G_\mathrm{RP})_0$ and $(G_\mathrm{RP}-W3)_0$ 
and absolute magnitudes $M_\mathrm{G_\mathrm{BP}}$, $M_\mathrm{G_\mathrm{RP}}$, and $M_\mathrm{W3}$ with $|Z|$.

Without an application of any model of dust distribution (or spatial variations of extinction/reddening), 
we analyzed in \citetalias{gm2020} the spatial variations of the observables within 
(i) a thin coordinate layer of $|Z|<40$~pc, 
(ii) a narrow vertical cylinder of $(X^2+Y^2)^{0.5}<80$~pc (where $X$ and $Y$ are the Galactic rectangular coordinates), and 
(iii) a space behind the layer limited by $|Z|>400$ pc and the latitude $|b|>68.75\degr$.
In particular, this allowed us to obtain the following estimates of the absolute magnitudes and intrinsic colours of the clump:
\begin{equation}
\label{abs1}
M_\mathrm{G_\mathrm{BP}}=(0.91\pm0.01)+(0.01\pm0.01)\,|Z|\,,
\end{equation}
\begin{equation}
\label{abs2}
M_\mathrm{G_\mathrm{RP}}=(-0.235\pm0.01)+(0.04\pm0.01)\,|Z|\,,
\end{equation}
\begin{equation}
\label{abs3}
M_\mathrm{W3}=(-1.67\pm0.01)+(0.07\pm0.01)\,|Z|\,,
\end{equation}
\begin{equation}
\label{color1}
(G_\mathrm{BP}-G_\mathrm{RP})_0=(1.145\pm0.01)-(0.03\pm0.01)\,|Z|\,,
\end{equation}
\begin{equation}
\label{color2}
(G_\mathrm{RP}-W3)_0=(1.435\pm0.01)-(0.03\pm0.01)\,|Z|\,,
\end{equation}
where $Z$ is in kpc.

We compared these values with the theoretical predictions from PAdova and TRieste Stellar Evolution Code 
(PARSEC, \citealt{marigo2017}\footnote{\url{http://stev.oapd.inaf.it/cgi-bin/cmd}}), MESA Isochrones and Stellar Tracks 
(MIST, \citealt{paxton2011, paxton2013, choi2016, mist}\footnote{\url{http://waps.cfa.harvard.edu/MIST/}}), and
the Bag of Stellar Tracks and Isochrones (IAC-BaSTI, \citealt{newbasti}\footnote{\url{http://basti-iac.oa-abruzzo.inaf.it/index.html}.
IAC-BaSTI does not provide any estimate for $W3$ band.}).
This allowed us to estimate the clump median age and [Fe/H] with their linear vertical gradients within $|Z|<1.7$ kpc as 
$(2.3\pm0.5)+(3.2\pm1.6)\,|Z|$ Gyr and $(-0.08\pm0.08)-(0.16\pm0.07)\,|Z|$, respectively, where $Z$ is expressed in kpc.

In \citetalias{gm2020} we found the extinctions and reddenings across the whole dust half-layer below or above the Sun. 
They converge to the reddening $E(B-V)\approx0.06$ mag by use of the extinction laws of 
\citet[][hereafter DIB14]{davenport2014} for the high latitudes,
\citet[][hereafter SMS16]{schlafly2016}, \citet[][hereafter WC19]{wang2019}, and 
\citet[][hereafter CCM89]{ccm89} with $R_\mathrm{V}\equiv A_\mathrm{V}/E(B-V)=3.1$.

The aim of this paper is to use the same sample, but with a 3D dust distribution model applied, in order to simultaneously 
derive some properties of the Galactic dust layer and clump giants embedded into or seen through this layer.

To choose a proper 3D dust distribution model we take into account the note of \citet{dame2001}:
`most of the major local molecular clouds appear to follow Gould's Belt,\ldots
the apparent disk of OB stars, gas, and dust surrounding the Sun and inclined $\approx20\degr$ to the Galactic plane'.
\citet{av}, following \citet{gould}, proposed to consider two dust layers intersecting near the Sun: 
the equatorial one and the layer in the Gould Belt.
They were described in the first \citep[][hereafter G09]{gould} and second \citep[][hereafter G12]{av} versions of 
the model of the 3D dust distribution. 
We revise this model in the current study.

Then we verify this revised version of the model (hereafter GM20) among other dust distribution models and maps\footnote{A model 
is a representation by some formulas, while a map is a representation in a tabular view.}
in their ability to correctly reproduce the modes of the colours $G_\mathrm{BP}-G_\mathrm{RP}$ and $G_\mathrm{RP}-W3$ for our sample.

We should take into account some results of the previous verifications of the most advanced models and maps by
\citet{g17, gm2017, gm2017big, gm2018, polarization}.
They have shown that:
\begin{enumerate}
\item 2D maps by \citet[][hereafter SFD98]{sfd} and \citet[][hereafter MF15]{2015ApJ...798...88M}, which provide only a 
cumulative reddening to infinity, are useless in describing the distribution of the local dust.
\item The same is true for any 3D follower of these maps. An example is the map of \citet[][hereafter DCL03]{drimmel}, 
based on the model of \citet{drispe}. It includes a description of a dust segment of the local arm based on some dust emission 
at an unknown distance taken from a 2D map.  However, the source of this emission seems to be too distant.
\citetalias{drimmel} note: `Ideally this region about the Sun should be described by a more detailed local model of the dust 
distribution.'
\item Any incompleteness of the stellar sample in use may cause a bias in the results, usually towards low reddening.
This may be a reason for some unreliable estimates with the map of \citet[][hereafter LVV14]{lallement2014}.
\item The best representations of the local dust distribution have been obtained using map of \citet[][hereafter G17]{g17}.
It is successful due to the use of unsaturated photometry for a complete sample of nearby stars.
\citetalias{av} has proven to be the best among the models due to a direct consideration of the Gould Belt as an 
important local dust container.
\citet{gm2017big, gm2018} concluded that \citetalias{av} deserves more attention and a further refinement.
This is fulfilled in the current study.
\end{enumerate}

This paper is organized as follows. 
In Sect.~\ref{modewholespace} we analyze the spatial variations of the colour modes in order to derive a general pattern of 
the 3D dust distribution and to constrain a proper geometry of the model.
In Sect.~\ref{models} we consider a simplest model and GM20 to fit the observed variations of the colour modes.
We compare several 3D models and maps in their ability to fit the data in Sect.~\ref{compar}.
Some notes on the reddening across the whole dust layer are given in Sect.~\ref{whole}.
We summarize our findings and state our conclusions in Sect.~\ref{conclusions}.

\section{The modes throughout the whole space}
\label{modewholespace}

Some information about the 3D dust distribution can be derived from $mode(G_\mathrm{BP}-G_\mathrm{RP})$ and 
$mode(G_\mathrm{RP}-W3)$ calculated in some cells located throughout the whole space cylinder under consideration.

As we note in Sect.~\ref{intro}, to obtain the modes, we round the observables up to 0.01 mag and find the tops of their 
histograms in each cell.
In the case of multimodal histogram, the lowest value is selected, since it is more probable for the unreddened or 
slightly reddened clump.

As discussed in \citetalias{gm2020}, space cells with 400 giants are optimal to calculate the modes.
This number allows us to keep a balance between precision and spatial resolution of the results.
In such a cell, the uncertainty of the modes is defined by natural fluctuations of the interstellar medium between us and stars 
in the cell. These fluctuations are followed by related fluctuations of the extinction, reddening, and observables.

Initially, we set 101\,810 cells with our stars as their centres.
For each cell the modes of the colours are calculated by use of the stars within a radius of 
\begin{equation}
\label{cellradius}
110/(\mathrm{e}^{-|Z|/1000}),
\end{equation} 
where $Z$ is expressed in pc. 
The radius varies from 110~pc at the Galactic mid-plane to 665~pc at $|Z|=1800$~pc. Yet, within the dust layer it 
increases slowly: 
to only 181 pc at $|Z|=500$~pc. Such radius allows us to use about 400 stars in each cell and calculate the colour 
modes with a precision at a level of 0.02 mag.
Finally, in order to take into account a non-uniform spatial distribution of the stars, 
the calculated modes are referred to average positions of the stars used in the cells instead of the centres of the cells.

For each cell we calculate a distance between the average position of the stars used and the centre of the cell.   
A small ratio of this distance to the cell radius may be due to a random deviation of the stars from their uniform 
spatial distribution.   A large ratio may mean a systematic loss of stars in a cell at the periphery of the space cylinder.
This may lead to a bias in the related results.
Therefore, we estimate the maximal acceptable ratio due to random reasons as a function of $|Z|$ by a Monte-Carlo simulation 
based on the spatial distribution of the stars. We use a million realizations of the simulated stars.
The desired ratio appears about 0.2 for $|Z|<700$ pc, increasing as $0.1/\exp(-|Z|)$ for higher $|Z|$, 
where $Z$ is in kpc, due to a lower spatial density of the stars.    We use these limits to cut out the biased cells.
The modes of the colours for the remaining 83\,317 cells are the main data for our further analysis.
These data are presented in Table~\ref{modes} available online.

\begin{table}
\def\baselinestretch{1}\normalsize\normalsize
\caption[]{The main data for 83\,317 spatial cells under consideration: $X$, $Y$ and $Z$ coordinates (pc), 
$mode(G_\mathrm{BP}-G_\mathrm{RP})$ (mag), $mode(G_\mathrm{RP}-W3)$ (mag). 
The complete table is available in the Supporting Information.
}
\label{modes}
\[
\begin{tabular}{rrrrr}
\hline
\noalign{\smallskip}
$X$ & $Y$ & $Z$ & $mode(G_\mathrm{BP}-G_\mathrm{RP})$ & $mode(G_\mathrm{RP}-W3)$  \\
\hline
\noalign{\smallskip}
   0 &  -1 &  165 & 1.19 & 1.53 \\
  -3 &  -2 &  253 & 1.20 & 1.58 \\
  -4 &  -1 &  179 & 1.19 & 1.58 \\
   1 &  -5 &   72 & 1.22 & 1.43 \\
   3 &  -5 &  339 & 1.20 & 1.58 \\
\ldots & \ldots & \ldots & \ldots & \ldots \\
\hline
\end{tabular}
\]
\end{table}

\begin{figure*}
\includegraphics{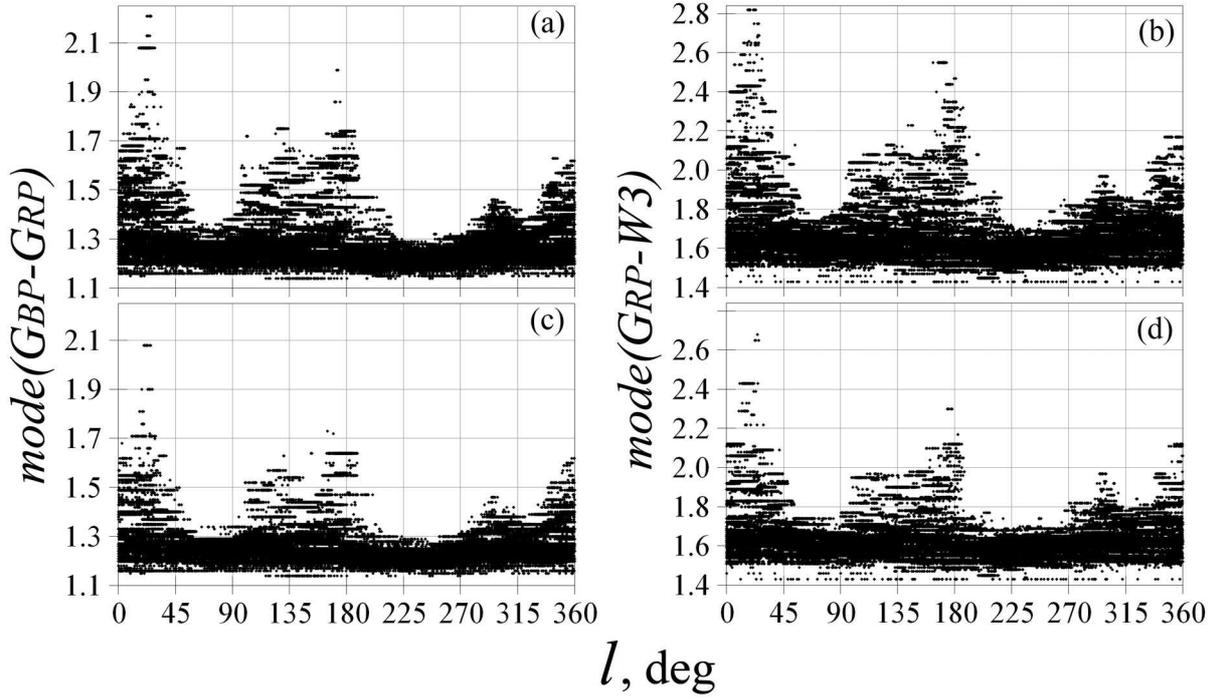}
\caption{(a) $mode(G_\mathrm{BP}-G_\mathrm{RP})$ and (b) $mode(G_\mathrm{RP}-W3)$ as functions of $l$ for the cylinder of $(X^2+Y^2)^{0.5}<700$~pc.
(c) and (d) -- the same for the cylinder of $(X^2+Y^2)^{0.5}<500$~pc.
}
\label{longi}
\end{figure*}

\begin{figure}
\includegraphics{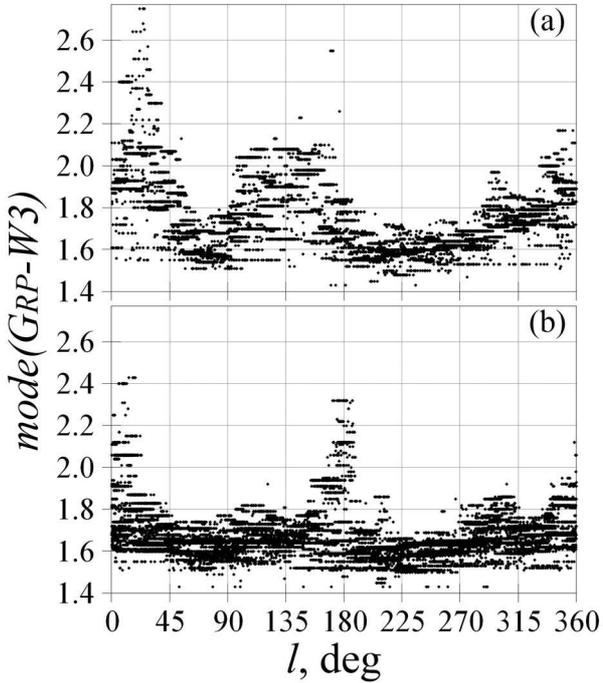}
\caption{$mode(G_\mathrm{RP}-W3)$ as a function of $l$ for (a) $|b|<5\degr$ and (b) $18\degr<|b|<30\degr$.
}
\label{lgrw3}
\end{figure}

\subsection{Longitudinal variations}
\label{longvar}

\begin{figure*}
\includegraphics{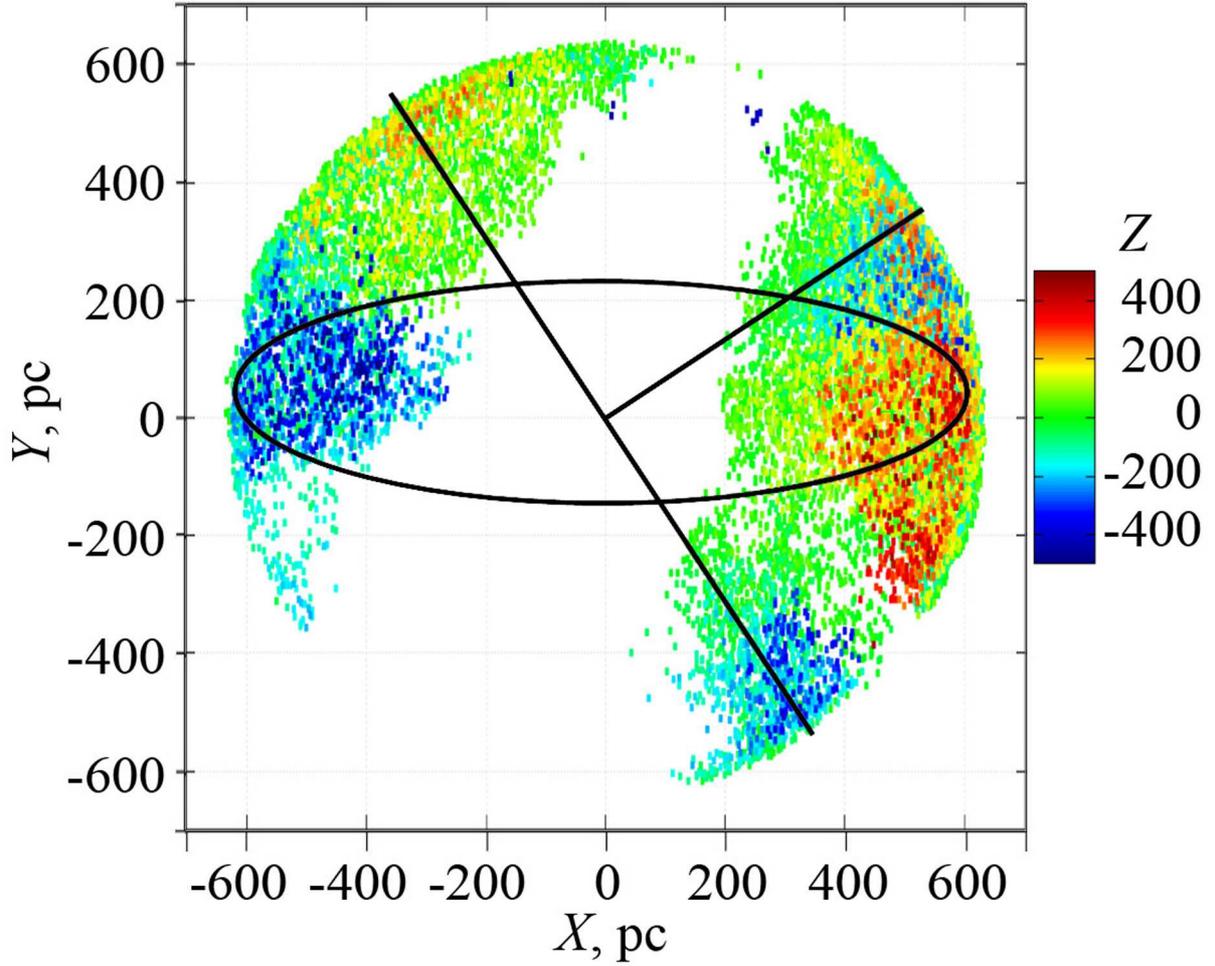}
\caption{The cells with $mode(G_\mathrm{BP}-G_\mathrm{RP})>1.35$ projected into the $XY$ plane with their $Z$ 
coordinate shown by the colour scale at the right. The Sun is at the centre. The Galactic centre is to the right.
The approximate position of the Gould Belt is shown by the black ellipse.
The directions to the three regions with a higher reddening in the equatorial dust layer are shown by the black lines.
}
\label{xy135}
\end{figure*}

Fig.~\ref{longi} shows the modes of the colours calculated in the 83\,317 cells as functions of 
the longitude $l$ for the cylinders of $(X^2+Y^2)^{0.5}<700$ and $(X^2+Y^2)^{0.5}<500$~pc.
The pattern in all the plots is similar and well known.
For example, it can be compared with figure 6 of \citet{gould} obtained for a different stellar class, OB-stars.
This pattern draws four distinct regions (longitudes ranges) with a higher reddening 
$mode(G_\mathrm{BP}-G_\mathrm{RP})>1.35$ and $mode(G_\mathrm{RP}-W3)>1.75$ at 
$l\approx5\degr$, $120\degr$, $175\degr$, and $300\degr$. The fifth region at $l\approx30\degr$ is less distinct.

For a better understanding of the pattern, Fig.~\ref{lgrw3} shows $mode(G_\mathrm{RP}-W3)$ as a function of $l$ 
for two ranges of latitude: (a) $|b|<5\degr$ and (b) $18\degr<|b|<30\degr$.
The equatorial dust layer should dominate in the former, while the Belt layer -- in the latter.
Indeed, Fig.~\ref{lgrw3} shows different patterns in the different ranges of $|b|$.
It attributes the three regions of a higher reddening to the equatorial and the two regions to the Belt layer.
$mode(G_\mathrm{BP}-G_\mathrm{RP})$ draws a similar pattern.

The regions at $l\approx30\degr$, $120\degr$, and $300\degr$ lie near the Galactic mid-plane and are separated 
from each other by $\Delta l\approx90\degr$.
These regions are seen as three petals of a similar colour (i.e. similar $Z$ coordinates) in Fig.~\ref{xy135}, 
where the cells 
with the high reddening $mode(G_\mathrm{BP}-G_\mathrm{RP})>1.35$ are projected into the $XY$ plane 
with their $Z$ coordinate shown by the colour, with its scale indicated at the right.
The directions from the Sun (at the centre) to these three higher reddening petals are shown by the black lines.
This pattern suggests some sinusoidal variations of the reddening in the equatorial dust layer with $l$ 
(with the maximal reddening at $l\approx30\degr$) 
and with $2\,l$ (with the maximal reddening at $l\approx120\degr$ and $300\degr$).
The cells with $mode(G_\mathrm{RP}-W3)>1.75$ show a similar pattern.
The sinusoid of $l$ may be explained by the well-known gradient of the dust medium density in the directions 
to the centre and anticentre of the Galaxy. The sinusoid of $2\,l$ may be related to the Local Arm bisecting across local space.

However, the colour of the petals at $l\approx30\degr$, $120\degr$, and $300\degr$ in Fig.~\ref{xy135} is not exactly the same. 
This points to a deviation of the petals from the Galactic mid-plane to the north and south in the second and fourth Galactic 
quadrants, respectively\footnote{The colour of the cells in the first quadrant in Fig.~\ref{xy135} suggests a deviation of its 
dust layer to the south.}.
This deviation is a warp of the equatorial dust layer.
This warp has been known before. For example, it is seen as a higher reddening at $-25\degr<b<0\degr$ in the fourth quadrant and 
$0\degr<b<+25\degr$ in the second quadrant in figure 6 of \citet{g2010}, in addition to an even higher reddening 
in the Gould Belt seen in the same figure at the directions to the centre and anticentre of the Galaxy.

The regions of these petals with the most deviation from the mid-plane correspond to the well-known Aquila South 
($l\approx33\degr$, $b\approx-17\degr$), Polaris Flare ($l\approx120\degr$, $b\approx+25\degr$), 
and Chamaeleon ($l\approx300\degr$, $b\approx-17\degr$) cloud complexes \citep[][figure 2a inset]{dame2001}.
Yet, despite the deviation of these complexes from the Galactic mid-plane, the petals under consideration are 
dominated by the dust near the mid-plane, as seen from the colours in Fig.~\ref{xy135}.

Finally, the 3D distribution of the cells in these three petals allows us to impose some constraints on the characteristics of 
the equatorial dust layer. They must be: the layer scale height $70<Z_\mathrm{A}<250$ pc, the phase $35\degr<\Phi<75\degr$ of 
the reddening variations with $\sin(l+\Phi)$ and the corresponding phase $\Phi'=2\Phi+90\degr$ of the reddening variations 
with $\sin(2l+\Phi')$.

The other two regions of higher reddening at $l\approx5\degr$ and $175\degr$ are seen in Fig.~\ref{xy135} as two petals far from 
the Galactic mid-plane.
Their deviation from the mid-plane, as follows from their colour in Fig.~\ref{xy135}, corresponds to $|Z|\approx180$ pc 
at $|X|\approx500$ pc and, consequently, a tilt of $\arctan(180/500)=20\degr$.
Thus, these petals follow the known orientation of the Gould Belt. Its mid-plane is inclined to the Galactic mid-plane by nearly 
$18\degr$ with the upper and lower sides directed nearly to the centre and anticentre of the Galaxy, respectively 
\citep{perryman} (pp. 311--314, 324--328).
Consequently, these two petals of higher reddening contain the Aquila rift, Ophiuchus, Lupus (at the Belt upper side) and 
Taurus -- Perseus -- Auriga, Orion (at its lower side) cloud complexes, respectively, 
as seen from \citet[][figure 2a inset]{dame2001}.

Note that the difference of the longitudes of the high reddening petals $\Delta l=175\degr-5\degr=170\degr$ deviates from $180\degr$.
This means that the centre of the Belt layer is shifted from the Sun to the second Galactic quadrant.

An approximate position of the Belt layer is drawn schematically in Fig.~\ref{xy135} by the ellipse.

Finally, the 3D distribution of the Belt cells allows us to put some constraints on the position, size, form and orientation of 
the Belt layer, assuming it is finite.
It can be described as an ellipse with the semi-major axis $400$~pc$<A<600$~pc, the semi-minor axis $85$~pc$<a<250$~pc, 
the eccentricity $0.78<e<0.99$, and the inclination to the Galactic mid-plane $15\degr<\gamma<21\degr$.
The Belt layer centre is shifted by $-100<x_0<0$, $+50<y_0<+150$, and $-50$~pc$<z_0<0$~pc along the $X$, $Y$, and $Z$ Galactic 
coordinate axes, respectively.
The Belt layer has the scale height $50<\zeta_\mathrm{A}<200$~pc, the angle $-20\degr<\eta<+20\degr$ between the semi-major axis 
and the direction of the maximum reddening, the angle $-20\degr<\theta<+20\degr$ between the $Y$ Galactic coordinate axis and 
the line of the intersection of the mid-planes of the equatorial and Belt layers, and the phase $70\degr<\phi<110\degr$ of the 
reddening variations with $\sin(2\lambda+\phi)$, where $\lambda$ is the longitude in the Belt mid-plane, which will be defined later.

The space with the lowest reddening, i.e. $mode(G_\mathrm{BP}-G_\mathrm{RP})<1.35$, between the five petals of the higher reddening,
is seen in Fig.~\ref{xy135} as a central white region.
It is elongated in Fig.~\ref{xy135} from the first to the third Galactic quadrant.
It is the well-known Local interstellar tunnel, or the Great tunnel \citep{welsh1991, av, polarization}.
Its appearance in Fig.~\ref{xy135} confirms our results.

Fig.~\ref{longi}~(a), (b) and Fig.~\ref{lgrw3}~(b) shows that the petals of the Belt are slightly asymmetric 
w.r.t. the Sun: the maximum reddening at $l\approx5\degr$ is higher than the one at $175\degr$ for both the colours.
It can be explained by the shift of the Sun w.r.t. the centre of the Belt layer.

A more pronounced asymmetry is seen for the petals of the equatorial layer, especially in Fig.~\ref{lgrw3}~(a):
the maximum reddening at $l\approx120\degr$ is higher than the one at $300\degr$ for both the colours.
It can be explained by an asymmetry of the equatorial layer around the Sun, namely, by an additional dust container in the second 
Galactic quadrant. Fig.~\ref{longi}~(d) shows that this asymmetry in the equatorial layers disappears for $mode(G_\mathrm{RP}-W3)$, 
if we limit the space by $(X^2+Y^2)^{0.5}<500$~pc: the maxima at $120\degr$ and $300\degr$ becomes equal.
However, this is not the case for $mode(G_\mathrm{BP}-G_\mathrm{RP})$, for which such an asymmetry persists within 
$(X^2+Y^2)^{0.5}<500$~pc, as seen in Fig.~\ref{longi}~(c), and disappears only within $(X^2+Y^2)^{0.5}<400$~pc. 
Such different spatial variations of the colour modes within $400<(X^2+Y^2)^{0.5}<500$~pc are possible, 
since their linear correlation coefficient in the whole space cylinder under consideration is only 0.86.
The different spatial variations of $mode(G_\mathrm{BP}-G_\mathrm{RP})$ and $mode(G_\mathrm{RP}-W3)$ mean that the variations 
of $E(G_\mathrm{BP}-G_\mathrm{RP})$ and $E(G_\mathrm{RP}-W3)$ differ.
In turn, this suggests a slightly different spatial distribution of fine and coarse dust grains, which are responsible for 
the former and latter reddening, respectively.

Thus, any model of the 3D dust distribution in the first kiloparsec should take into account 
(i) the additional dust container in the second quadrant beyond 400 pc from the Sun and 
(ii) possible slightly different spatial distribution of the fine and coarse dust and related possible spatial variation of the 
extinction law near the Galactic mid-plane.
However, GM20 presented in this study, does not take into account any warp or asymmetry of the equatorial dust layer.
Moreover, up to now there is no model which takes these features into account correctly.
This can potentially be a subject for future studies.

\section{Models}
\label{models}

\subsection{Simplest model}
\label{simplest}

The simplest model implies an exponential vertical distribution of dust in one layer, without any variation on $l$.
Hence, the cumulative reddening of a star follows the barometric law \citep[][ p. 265]{parenago}:
\begin{equation}
\label{singlelayer}
E_0\,R(1-\mathrm{e}^{-|Z-Z_\mathrm{0}|/Z_\mathrm{A}})Z_\mathrm{A}/|Z-Z_\mathrm{0}|\,,
\end{equation}
where $E_0$ is the differential reddening in the mid-plane of the layer, $R$ is the distance from the Sun,
$Z_\mathrm{0}$ is the vertical offset for the mid-plane of the dust layer with respect to the Sun, and
$Z_\mathrm{A}$ is the scale height of the dust layer.

Also, we derive one additional parameter for each colour -- the linear gradient of the dereddened colour 
$(G_\mathrm{BP}-G_\mathrm{RP})_0$ or $(G_\mathrm{RP}-W3)_0$ with $|Z|$.
Note that these colours themselves can be derived for the simplest model only with a very large uncertainty. 
Hence, they are fixed as 1.145 and 1.435, respectively, following the findings of \citetalias{gm2020}.
We derive the least root-mean-square deviations of this model from the data in the 83\,317 cells with the following parameters:
$Z_0=-6$ pc for $mode(G_\mathrm{BP}-G_\mathrm{RP})$ versus $-1$~pc for $mode(G_\mathrm{RP}-W3)$,
$Z_\mathrm{A}=170$ versus $180$~pc, $E_0=0.4$ versus $0.5$~mag~$kpc^{-1}$,
$\Delta (G_\mathrm{BP}-G_\mathrm{RP})_0=-0.024|Z|$ versus $\Delta (G_\mathrm{RP}-W3)_0=-0.018|Z|$, where $Z$ is in kpc.
We note the rather large scale heights as compared to estimates in the literature \citep[][pp. 470--471, 496--497]{perryman}.
Yet, they are rather diverse: from $35$ \citep{vergely1998} and $<70$ \citep{juric} to $140$ \citep{bmg1} and $188$~pc \citep{drispe}.

The derived reddenings across the whole dust half-layer below or above the Sun are:
\begin{equation}
\label{simplegbgr}
E(G_\mathrm{BP}-G_\mathrm{RP})=0.068\,,
\end{equation}
\begin{equation}
\label{simplegrw3}
E(G_\mathrm{RP}-W3)=0.090\,.
\end{equation}
By use of the extinction law of \citetalias{wang2019} the values (\ref{simplegbgr}) and (\ref{simplegrw3}) correspond 
to $E(B-V)=0.051$ and 0.052 mag, respectively. 
These estimates will be compared with others in Sect.~\ref{whole}.

The corresponding colour excess ratio (CER) is $E(G_\mathrm{RP}-W3)/E(G_\mathrm{BP}-G_\mathrm{RP})=1.32$.
It fits good with the CERs 1.24, 1.28, and 1.33 provided by the extinction laws of \citetalias{davenport2014} for the high latitudes,
\citetalias{schlafly2016}, and \citetalias{wang2019}, respectively.
However, our CER disagrees with the CER 1.69 from the law of \citetalias{ccm89} with $R_\mathrm{V}=3.1$.

\subsection{Model GM20}
\label{model}

\begin{figure*}
\includegraphics{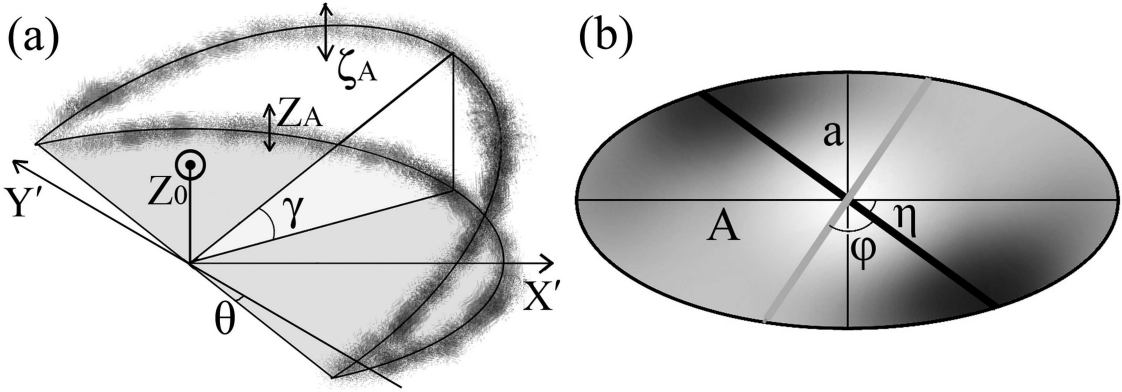}
\caption{The scheme of the mid-planes of the Galactic dust layers near the Sun:
(a) for the two intersecting mid-planes, (b) for the elliptical mid-plane of the Gould Belt.
$X'$ and $Y'$ are the projections of the Galactic coordinate axes $X$ and $Y$ into the mid-plane of the equatorial dust layer.
}
\label{g19}
\end{figure*}

Here we apply the model GM20 to the data in order to derive a set of its parameters together with the additional parameters -- 
the modes of the dereddened colour $(G_\mathrm{BP}-G_\mathrm{RP})_0$ or $(G_\mathrm{RP}-W3)_0$ and their linear gradients with $|Z|$.

GM20 implies the existence of two dust layers with their mid-planes intersecting at an angle $\gamma$:
the equatorial layer and the layer in the Gould Belt, with the scale heights $Z_\mathrm{A}$ and $\zeta_\mathrm{A}$, respectively.
The line of the intersection of the mid-planes of these layers is rotated from the $Y$ axis by the angle $\theta$.
The scheme of the mid-planes of the layers is shown in Fig.~\ref{g19}. Note that we do not consider any warp of the layers.

Expanding far beyond the space under consideration, the equatorial layer is considered as infinite along the $X$ and $Y$ coordinates.
The equatorial layer mid-plane is shifted  w.r.t. the Sun by $Z_\mathrm{0}$ along the $Z$ coordinate.

In contrast, the Gould Belt layer mid-plane is limited by the ellipse with the semi-major axis $A$, semi-minor axis $a$, 
eccentricity $e$, and angle $\eta$ between the semi-major axis and the direction of the maximum reddening.
The centre of this ellipse is shifted w.r.t. the Sun by $x_\mathrm{0}$, $y_\mathrm{0}$, and $z_\mathrm{0}$ along $X$, $Y$, and $Z$, 
respectively.
This is a difference of GM20 from its previous versions, where the Belt mid-plane was limited by a circle shifted w.r.t. the Sun 
only along the direction perpendicular to the Belt mid-plane.

Another difference of GM20 from its previous versions follows from the findings in Sect.~\ref{longvar}:
the longitudinal variations of reddening in the equatorial layer include two terms, with $\sin(l)$ and $\sin(2l)$.
Therefore, the cumulative reddening $E(G_\mathrm{BP}-G_\mathrm{RP})$ or $E(G_\mathrm{RP}-W3)$ is the sum of three terms.
They follow a barometric law and vary along the mid-plane of the corresponding layer: 
\begin{equation}
\label{equ1}
[E+S\sin(l+\Phi)]R(1-\mathrm{e}^{-|Z-Z_\mathrm{0}|/Z_\mathrm{A}})Z_\mathrm{A}/|Z-Z_\mathrm{0}|\,,
\end{equation}
\begin{equation}
\label{equ2}
[D\sin(2l+2\Phi+90\degr)]R(1-\mathrm{e}^{-|Z-Z_\mathrm{0}|/Z_\mathrm{A}})Z_\mathrm{A}/|Z-Z_\mathrm{0}|
\end{equation}
for the single and double sinusoidal longitudinal variations in the equatorial layer, respectively, and
\begin{equation}
\label{equ3}
[G_0+G_1\sin(2\lambda+\phi)]\min(R,R_0)(1-\mathrm{e}^{-|\zeta|/\zeta_\mathrm{A}})\zeta_\mathrm{A}/|\zeta|
\end{equation}
for the variations in the Belt layer. $E$ and $G_0$ are the constant terms, $S$, $D$, and $G_1$ are the amplitudes, 
and $\Phi$, $2\Phi+90\degr$ and $\phi$ are the phases.
The coordinates of a star/point/cell in the coordinate system of the Belt are the latitude $\beta$ counted from the Belt mid-plane, 
the longitude $\lambda$ counted from the direction of the maximal reddening, and the distance from the mid-plane of the Belt $\zeta$.
$R_0$ is the distance from the Sun to the Belt layer edge at the line of sight of a star under consideration.
The term $\min(R,R_0)$ in equation~(\ref{equ3}) allows us to calculate the reddening in the Belt layer to its edge or to the star, 
whichever is closer.
These quantities are calculated as: 
\begin{equation}
\label{r0}
R_0=a/(1-[\mathrm{e}\cos(\lambda-\eta)]^2)^{0.5}\,,
\end{equation}
\begin{equation}
\label{equbeta}
\sin(\beta)=\cos(\gamma)\sin(b)-\sin(\gamma)\cos(b)\cos(l)\,,
\end{equation}
\begin{equation}
\label{equlambda}
\tan(\lambda-\theta)=\cos(b)\sin(l)/[\sin(\gamma)\sin(b)+\cos(\gamma)\cos(b)\cos(l)]\,,
\end{equation}
\begin{equation}
\label{equzeta}
\zeta=\min(R,R_0)\sin(\beta)
\end{equation}

We assume that $Z_\mathrm{0}$ and $Z_\mathrm{A}$ are the same for the terms of the equatorial layer.
This assumption is based on our analysis of Fig.~\ref{xy135} and its analogues in the $XZ$ and $YZ$ planes.
This assumption wiil be fully validated in our future work, where these parameters will be made free.

\subsection{GM20 results}
\label{results}

\begin{table*}
 \centering
\def\baselinestretch{1}\normalsize\normalsize
\caption[]{The parameters of the Galactic dust layer and the giant clump in GM20.
}
\label{solution}
\begin{tabular}[c]{lccc}
\hline
\noalign{\smallskip}
 Parameter                                 & Constraints & $G_\mathrm{BP}-G_\mathrm{RP}$ & $G_\mathrm{RP}-W3$ \\
\hline
\noalign{\smallskip}
Dereddened colour (mag)                     & $1.10-1.18$ / $1.41-1.52$   & $1.14\pm0.01$     & $1.44\pm0.01$  \\
Colour gradient along $|Z|$ (mag kpc$^{-1}$)   & $-0.08-0$                   & $-0.022\pm0.010$  & $-0.015\pm0.010$ \\
$\gamma$ ($\degr$)                          & $15-21$                     & $17\pm2$          & $17\pm2$       \\
$\theta$ ($\degr$)                          & $-20-+20$                   & $-3\pm4$          & $-4\pm4$       \\ 
$\eta$ ($\degr$)                            & $-20-+20$                   & $2\pm4$           & $0\pm4$       \\ 
$Z_0$ (pc)                                  & $-30-+30$                   & $-7\pm10$         & $-10\pm10$       \\ 
$x_0$ (pc)                                  & $-100-0$                    & $-25\pm15$        & $-5\pm15$       \\ 
$y_0$ (pc)                                  & $+50-+150$                    & $+135\pm15$       & $+115\pm15$       \\ 
$z_0$ (pc)                                  & $-50-0$                     & $-26\pm15$        & $-30\pm15$       \\ 
$A$ (pc)                                    & $400-650$                   & $600\pm50$        & $600\pm50$     \\ 
$a$                                        &                             & $146\pm37$        & $146\pm37$      \\
$e$                                        & $0.78-0.99$                 & $0.97\pm0.01$     & $0.97\pm0.01$     \\ 
$Z_A$ (pc)                                  & $70-250$                    & $150\pm15$        & $180\pm15$      \\ 
$\zeta_A$ (pc)                              & $50-200$                    & $55\pm15$         & $65\pm15$      \\ 
$E$ (mag kpc$^{-1}$)                           & $<1$                        & $0.35\pm0.07$     & $0.46\pm0.10$  \\ 
$S$ (mag kpc$^{-1}$)                           & $<1$                        & $0.04\pm0.07$     & $0.04\pm0.10$  \\ 
$\Phi$ ($\degr$)                            & $35-75$                     & $39\pm10$          & $39\pm10$       \\ 
$D$ (mag kpc$^{-1}$)                           & $<1$                        & $0.16\pm0.07$     & $0.20\pm0.10$  \\ 
$G0$ (mag kpc$^{-1}$)                          & $<1$                        & $0.45\pm0.07$     & $0.50\pm0.10$  \\ 
$G1$ (mag kpc$^{-1}$)                          & $<1$                        & $0.22\pm0.07$     & $0.24\pm0.10$  \\
$\phi$ ($\degr$)                            & $70-110$                    & $98\pm10$          & $98\pm10$       \\
\hline
\end{tabular}

\end{table*}


The solution, obtained for the 20 free parameters of GM20, is presented in Table~\ref{solution} (only two quantities among $A$, 
$a$ and $e$ are independent).
This solution corresponds to the least root-mean-square deviation of GM20 from the data [$mode(G_\mathrm{BP}-G_\mathrm{RP})$ and 
$mode(G_\mathrm{RP}-W3)$ in the 83\,317 cells]. 
This solution was achieved through a consideration of several trillions sets of the parameters varied within the constraints, listed
in Table~\ref{solution}, and the direct calculation of the residuals and their root-mean-square deviation.

Since one star is used for the mode calculation in several cells, the data in the cells are not independent.
To check the robustness of the solution and to estimate the uncertainties of the derived model parameters, we apply 
a resampling method.
We fill our cylinder space by 122 spherical cells, positioned so that they adjoin but do not intersect each other.
The size of each cell is defined by the $Z$ coordinate of its centre following relation (\ref{cellradius}).
For example, the mid-plane layer cells have a radius of 110 pc. The set of the cells does not fill the whole space cylinder.
We shift the set as a whole by 25 pc steps within $-50<X<50$ or $-50<Y<50$ or by 10 pc steps within $-10<Z<10$ pc to produce 
$5 \times 5 \times 3=75$ samples of the cells inside the space cylinder.
This allows us to use almost all the stars of the cylinder.
For each sample we calculate the colour modes for the central coordinates of each cell by use of all the stars in that cell.
Since the cells do not intersect each other, the mode values are independent of each other.
These modes are used to obtain a solution, with 20 parameters of our model, giving the least root-mean-square deviation of the 
model from the data.
Thus, one sample of the cells gives us one realization of the resampling method and, consequently, one set of the derived parameters.
The final distribution of each parameter in the 75 realizations/solutions appears to be nearly Gaussian.
The average values of the parameters are close to those for 83\,317 cells in Table~\ref{solution}.
The standard deviations of the parameters in 75 realizations/solutions are presented in Table~\ref{solution} as the uncertainties 
of the parameters.

The solutions for the two colours are the same within the uncertainties, except their $Z_\mathrm{A}$, and, naturally, 
dereddened colours.
Yet, the difference of $2\sigma$ between $Z_\mathrm{A}$ for two colours is not significant.
It is worth noting that all the derived angles agree with the position of the five petals of higher reddening in Fig.~\ref{xy135}.

The reddening maxima of the Belt layer in Fig.~\ref{longi}, \ref{lgrw3} and \ref{xy135} appear nearly at the directions to the 
Galactic centre and anticentre.
The eccentricity (0.97) of the Belt layer mid-plane is rather large. The differences $G_0-G_1$ is rather moderate.
All these facts mean that our model successfully reproduces a significant fraction of the Belt dust, which is contained in two, 
nearly opposite, regions at $-30\degr<l<+30\degr$, $+5\degr<b<+25\degr$ and $150\degr<l<220\degr$ , $-25\degr<b<-5\degr$.
As pointed out in Sect.~\ref{longvar}, these regions include the Aquila rift, Ophiuchus, Lupus and 
Taurus -- Perseus -- Auriga, Orion cloud complexes, respectively.

The spatial distribution of the Belt dust can differ from the distribution of its gas, stellar clusters and associations. 
However, Table~\ref{solution} shows the position of the Belt layer centre w.r.t. the Sun (average $X=-15$, $Y=+125$, $Z=-28$, 
$R=129$~pc) near the position (albeit rather uncertain) derived earlier from the distribution of young stars 
\citep{bobylev2014}, see also \citet[][p. 325]{perryman}.

The scale heights $Z_\mathrm{A}$ and $\zeta_\mathrm{A}$ of the equatorial and Belt layers, respectively, show the ratio of 
$2.7-2.8$ in agreement with the note from \citet[][p. 325]{perryman}
`the Gould Belt is three times as compressed vertically as the Galactic belt, each shows the same increasing concentration for stars,
interstellar dust, and stellar groups'.

Our solution gives rather large scale heights $Z_\mathrm{A}$, which are comparable with those obtained for the simplest model 
and those of \citet{bmg1} and \citet{drispe} mentioned in Sect.~\ref{simplest}.
Regardless, our solution is preferable as it is based on a sample covering a wider range of $Z$ and, for the first time, 
definitely covering the entire depth of the dust layer (or layers).
Do note that such large scale heights are ultimately defined by the fact that the colour modes of our sample still increase (become 
redder) at a high $|Z|$ (say, 450 pc).
Some previous studies could not show a reddening behavior at such a high $|Z|$ due to poor or no data.
Some examples of such studies are \citet{vergely1998} with their very low estimate of the dust layer scale height (35 pc) and the 
3D reddening maps of \citetalias{lallement2014} and \citet[][hereafter LVV19]{lallement2019}\footnote{\url{http://stilism.obspm.fr}}.
It is also worth noting that a rather thick dust layer near the Sun in GM20 agrees with the scale height estimates (always $>120$~pc) 
of edge-on galaxies from \citet{degeyter2014} and \citet{mosenkov2018}, see their tables 3 and 4, respectively.

GM20 gives reddenings across the whole dust half-layer below or above the Sun:
\begin{equation}
\label{gm20gbgr}
E(G_\mathrm{BP}-G_\mathrm{RP})=0.073\,,
\end{equation}
\begin{equation}
\label{gm20grw3}
E(G_\mathrm{RP}-W3)=0.105\,,
\end{equation}
with nearly the same contribution of the equatorial and Belt layers: 0.035 versus 0.038 and 0.055 versus 0.050 mag to the former 
and latter reddening, respectively. These estimates will be compared with others in Sect.~\ref{whole}.

Estimates (\ref{gm20gbgr}) and (\ref{gm20grw3}) give the CER $1.44$, which fits well with the CERs 1.24, 1.28, and 1.33, 
provided by the extinction laws of \citetalias{davenport2014} for the high latitudes, \citetalias{schlafly2016}, and 
\citetalias{wang2019}, respectively, while fits slightly worse with 1.69 from the law of \citetalias{ccm89} with $R_\mathrm{V}=3.1$.

\subsection{HR diagrams and isochrones}
\label{hr}

\begin{figure*}
\includegraphics{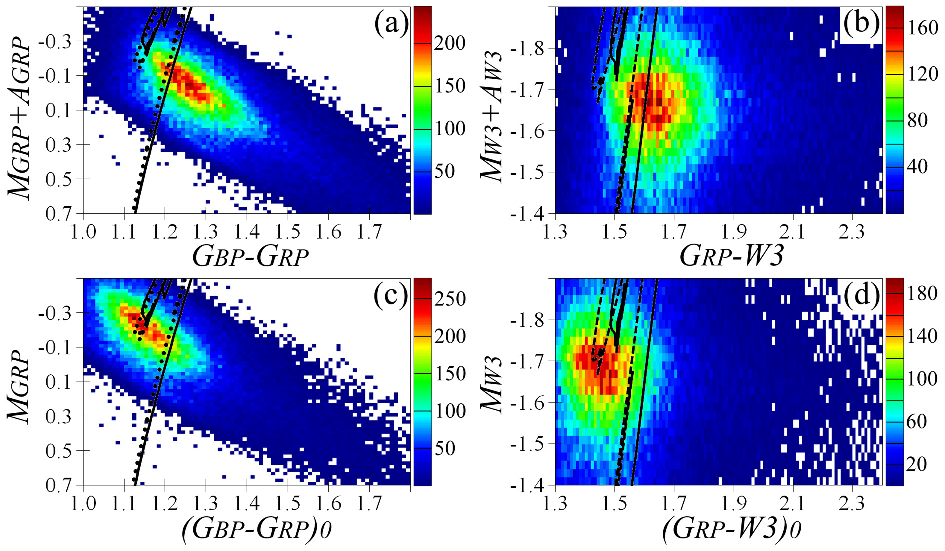}
\caption{The diagrams 
(a) `$G_\mathrm{BP}-G_\mathrm{RP}$ versus $M_\mathrm{G_\mathrm{RP}}+A_\mathrm{G_\mathrm{RP}}$',
(b) `$G_\mathrm{RP}-W3$ versus $M_\mathrm{W3}+A_\mathrm{W3}$',
(c) `($G_\mathrm{BP}-G_\mathrm{RP})_0$ versus $M_\mathrm{G_\mathrm{RP}}$', and
(d) `$(G_\mathrm{RP}-W3)_0$ versus $M_\mathrm{W3}$'
for the 101\,810 stars of our sample before and after the correction for the reddenings from GM20
in combination with the extinction law of \citetalias{wang2019}.
The bins are 0.01 and 0.02 mag for the abscissas and ordinates, respectively.
The number of the stars in each bin is shown by the colour scale on the right.
The PARSEC, MIST, and IAC-BaSTI isochrones for 3 Gyr and $\mathrm{[Fe/H]}=-0.11$ are shown by the dashed, solid and dotted black 
curves, respectively.
Since the PARSEC and MIST isochrones coincide with each other on the scale of plots (a) and (c),
the PARSEC isochrone is not shown in these plots.
The IAC-BaSTI presents no isochrone for plots (b) and (d).
}
\label{hr0}
\end{figure*}

All 101\,810 giants of our sample are shown in Fig.~\ref{hr0} with the HR diagrams before and after correcting for the reddenings 
$E(G_\mathrm{BP}-G_\mathrm{RP})$ and $E(G_\mathrm{RP}-W3)$ and extinctions $A_\mathrm{G_\mathrm{RP}}$ and $A_\mathrm{W3}$. 
The reddenings are derived in the current study with GM20, while the extinctions are calculated from the reddenings by use of 
the extinction law of \citetalias{wang2019}. It gives $A_\mathrm{G_\mathrm{RP}}=1.429E(G_\mathrm{BP}-G_\mathrm{RP})$ and 
$A_\mathrm{W3}=0.073E(G_\mathrm{RP}-W3)$.

Note that the extinction law is important in Fig.~\ref{hr0} only for the ordinates. 
Therefore, one should pay more attention to the distributions along the abscissas as our direct results.

For the median $|Z|$ of our sample (212 pc), \citetalias{gm2020} gives the median age 3 Gyr and $\mathrm{[Fe/H]}=-0.11$.
These estimates seem to be reasonable being in good agreement with the estimates from \citet{girardi2016}.
We show the corresponding PARSEC\footnote{PARSEC version 1.2S assuming $\mathrm{[\alpha/Fe]}=0$, solar metallicity $Z=0.0152$, 
mass-loss efficiency $\eta=0.2$, where $\eta$ is the free parameter in the Reimers' law \citep{reimers}.},
MIST\footnote{MIST version 1.2 assuming $\mathrm{[\alpha/Fe]}=0$ and the reference protosolar metallicity $Z=0.0142$.} 
and IAC-BaSTI\footnote{IAC-BaSTI version assuming $\mathrm{[\alpha/Fe]}=0$, overshooting, diffusion, mass-loss efficiency $\eta=0.3$.}
isochrones in Fig.~\ref{hr0} against the background of the distribution for the sample.
The clump median absolute magnitudes and intrinsic colours correspond to the bottom of the leftmost U shape loop of the isochrones.
The rightmost nearly vertical lines of the isochrones are the giant branch.

Comparing the original diagrams in plots (a) and (b) with the diagrams corrected for the reddening and extinction in plots (c) and 
(d), one can see that GM20 significantly reduces the scatter of the sample and suggests proper corrections.
Namely, the leftmost U shape loops of the isochrones fit the distribution tops within a few hundredths of a magnitude 
for the PARSEC, MIST and IAC-BaSTI isochrones in plot (c) and for the PARSEC isochrone in plot (d).
Our findings in \citetalias{gm2020} suggest that MIST is less reliable in its predictions for $W3$, probably,
due to an issue in its colour--$T_\mathrm{eff}$ relation and/or bolometric correction used.
A deviation of the MIST isochrone from the distribution top in plot (d) agrees with this suggestion.

Fig.~\ref{hr0} shows the distribution tops in perfect agreement with estimates (\ref{abs2})--(\ref{color2}) for the clump modes 
from \citetalias{gm2020}.

The distribution of the sample in Fig.~\ref{hr0} confirms that the branch and asymptotic branch giants are only a minor admixture 
to the clump giants, which does not affect the main modes of the distribution.
However, an admixture of the branch giants is noticeable as an excess of the stars just below the bulk in Fig.~\ref{hr0} (c) and (d), 
i.e. for
\begin{equation}
\label{rgb1}
(G_\mathrm{BP}-G_\mathrm{RP})_0\approx1.2
\end{equation}
and $M_\mathrm{G_\mathrm{RP}}\approx0.1$ or
\begin{equation}
\label{rgb2}
(G_\mathrm{RP}-W3)_0\approx1.5
\end{equation}
and $M_\mathrm{W3}>-1.5$ mag. 
It is seen that in plots (c) and (d) all the theoretical giant branches perfectly fit the observed giant branches, except the MIST 
giant branch in `$(G_\mathrm{RP}-W3)_0$ versus $M_\mathrm{W3}$'. In contrast, this excess of the branch giants is seen in
plots (a) and (b) at $G_\mathrm{BP}-G_\mathrm{RP}\approx1.3$ and $(G_\mathrm{RP}-W3)\approx1.66$.
The noticeable differences between these estimates and those (\ref{rgb1}) and (\ref{rgb2}) point out that GM20 accurately takes 
into account the considerable reddening of the sample.

The estimates from GM20:
\begin{equation}
\label{bprpresult}
(G_\mathrm{BP}-G_\mathrm{RP})_0=(1.14\pm0.01)-(0.022\pm0.010)\,|Z|\,,
\end{equation}
\begin{equation}
\label{rpw3result}
(G_\mathrm{RP}-W3)_0=(1.44\pm0.01)-(0.015\pm0.010)\,|Z|\,,
\end{equation}
where $Z$ is in kpc. 
They agree with estimates (\ref{color1}) and (\ref{color2}) obtained in \citetalias{gm2020}.

\section{Data versus models and maps}
\label{compar}

\begin{figure*}
\includegraphics{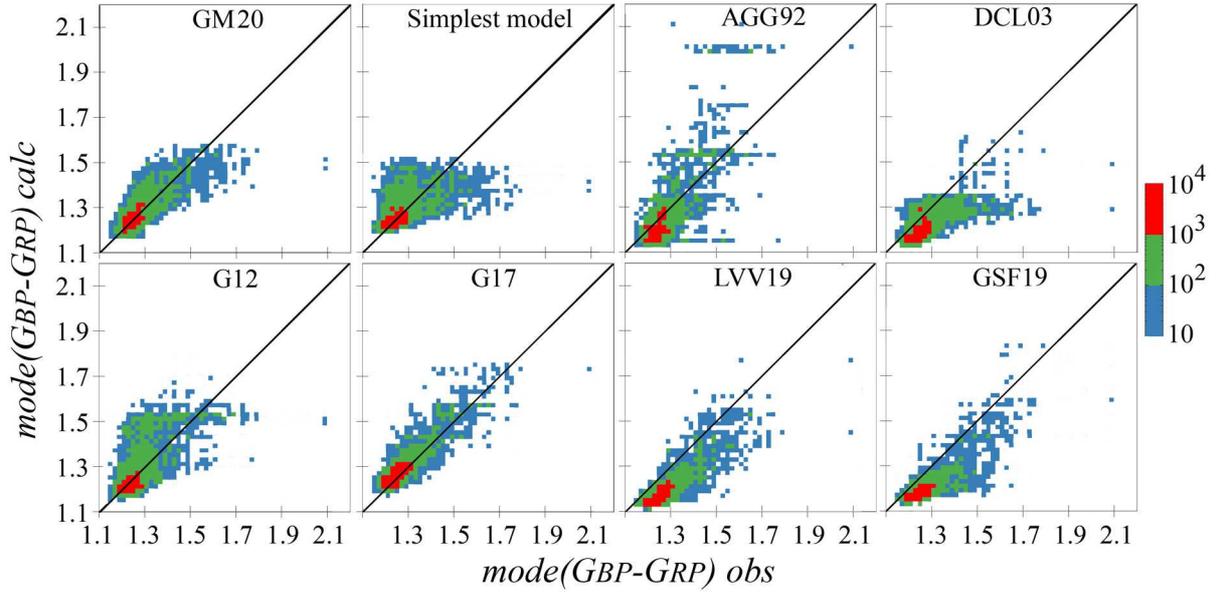}
\caption{The relation between the observed and calculated $mode(G_\mathrm{BP}-G_\mathrm{RP})$ for various models and maps.
The number of the stars in each bin of $0.02 \times 0.02$ mag is shown by the colour scale on the right.
}
\label{gbgrmode}
\end{figure*}

In Fig.~\ref{gbgrmode} we compare observed $mode(G_\mathrm{BP}-G_\mathrm{RP})$ with predictions from the most reliable 3D reddening 
models and maps\footnote{Some of these maps and models have been briefly described and compared with other data by 
\citet{gm2017big, gm2018}.}
for all 83\,317 cells.
$mode(G_\mathrm{RP}-W3)$ shows similar results. 
To convert the calculated reddenings into the colours, we use the extinction law of \citetalias{wang2019} and our 
estimates~(\ref{bprpresult}) and (\ref{rpw3result}).

Any reasonable change of the extinction law has little, if any, effect on this conversion.
The same is true for any reasonable change of the coefficients before $|Z|$ ($|Z|$ gradients) in estimates~(\ref{bprpresult}) and 
(\ref{rpw3result}), since these coefficients are small. 
With most stars of the sample within $|Z|<400$ pc, any reasonable change of these coefficients lead to a change of the dereddened 
colours~(\ref{bprpresult}) and (\ref{rpw3result}) of $<0.01$ mag.
Therefore, the only values, which can affect the comparison and, particularly, Fig.~\ref{gbgrmode}, are the dereddened 
colours~(\ref{bprpresult}) and (\ref{rpw3result}) at $Z=0$.
Any change of them leads to an overall offset of the model/map predictions, which would be seen as an overall vertical shift of 
the points in Fig.~\ref{gbgrmode}.

It is seen that such a vertical shift would give a better fit to the data for the predictions of \citetalias{drimmel}
\footnote{The \citetalias{drimmel} estimates are calculated by use of the code of \citet{bovy2016}, 
\url{https://github.com/jobovy/mwdust}}, \citetalias{lallement2019}, and 
\citet[][hereafter GSF19]{green2019}\footnote{\url{http://argonaut.skymaps.info/}}, 
i.e. would put the most amount of their points closer to the one--to--one relation shown by the black bisector line in 
Fig.~\ref{gbgrmode}.
More precisely, the best fit of their predictions to the data suggests
$(G_\mathrm{BP}-G_\mathrm{RP})_0=1.17$, 1.18 and 1.17; $(G_\mathrm{RP}-W3)_0=1.50$, 1.51 and 1.50 at $Z=0$, respectively.
However, these values are far from all the isochrone predictions in Fig.~\ref{hr0}, except the MIST isochrone for 
`$(G_\mathrm{RP}-W3)_0$ versus $M_\mathrm{W3}$'. Yet, the latter seems to be less reliable, as noted above.
Moreover, such high estimates of the colours cannot explain the existence of some cells with $mode(G_\mathrm{BP}-G_\mathrm{RP})<1.17$ 
and $mode(G_\mathrm{RP}-W3)<1.50$ seen, for example, in Fig.~\ref{longi}.
This may mean a systematic underestimation of the reddening by \citetalias{drimmel}, \citetalias{lallement2019}, 
and \citetalias{green2019}.
Such an underestimation, if real, introduces an especially large systematic uncertainty into low reddenings, 
i.e. for nearby and high-latitude stars.

It is seen in Fig.~\ref{gbgrmode} that the model of \citet[][hereafter AGG92]{arenou} slightly overestimates high and 
underestimates low reddenings.
This underestimation is explained by the constraint applied by \citetalias{arenou}: $A_\mathrm{V}<0.1$~mag for $|b|\ge60\degr$.

Also Fig.~\ref{gbgrmode} shows that GM20, \citetalias{av}, \citetalias{g17} and the simplest model from Sect.~\ref{simplest} drop 
the most cells near the bisector.
Consequently, these maps and models agree with estimates (\ref{bprpresult}) and (\ref{rpw3result}) (for GM20 -- by definition).

Such predictions by a perfect map/model should cover exactly the full range of $mode(G_\mathrm{BP}-G_\mathrm{RP})$ 
in Fig.~\ref{gbgrmode} with a minimal scatter around the bisector line.
It is worth noting that both the proper coverage of the range and the scatter are independent from estimates (\ref{bprpresult}) 
and (\ref{rpw3result}). The $|Z|$ gradients of the intrinsic colours in estimates (\ref{color1}), (\ref{color2}), 
(\ref{bprpresult}), and (\ref{rpw3result}) are negative. This makes distant stars bluer and having no effect on the nearby stars. 
Therefore, such a $|Z|$ gradient makes the predicted colour range narrower and does not shift the whole bulk of the points, 
i.e. does not change an offset.
Any lowering of the reddening also makes the predicted colour range narrower.
But no map/model needs to make its range narrower, as seen from Fig.~\ref{gbgrmode}. 
Moreover, we need wider colour ranges for the simplest model and \citetalias{drimmel}.

A rather moderate fit of the simplest model is achieved by the adjustment of its parameters in Sect.~\ref{simplest}.
The imperfection of this model confirms that the only dust layer without longitudinal variations, in any case, is not enough 
to describe the real 3D distribution of dust.

Fig.~\ref{gbgrmode} shows that GM20 is better than \citetalias{av}.
However, GM20 underestimates the highest reddenings, i.e. for $R>400$~pc, because it does not take into account the asymmetry of 
the equatorial layer, which was discussed in Sect.~\ref{longvar}.

Also Fig.~\ref{gbgrmode} shows that \citetalias{g17}, \citetalias{lallement2019}, and \citetalias{green2019} give nearly equally 
good predictions, if one pays no attention to the overall offsets, which depend on accepted estimates (\ref{bprpresult}) and 
(\ref{rpw3result}).

An increase of $(G_\mathrm{BP}-G_\mathrm{RP})_0$ and $(G_\mathrm{RP}-W3)_0$ at $Z=0$ w.r.t. estimates (\ref{bprpresult}) and 
(\ref{rpw3result}) can improve the overall offsets of \citetalias{drimmel}, \citetalias{lallement2019}, and \citetalias{green2019} 
from the data.
However, it cannot change the linear correlation coefficients between these predictions and the data.
Therefore, these correlation coefficients are useful to check, which prediction is better among those with a minimal scatter 
around the bisector line and with a proper coverage of the colour range.
Table~\ref{corr} with the correlation coefficients shows that, as expected from Fig.~\ref{gbgrmode}, \citetalias{g17}, 
\citetalias{lallement2019}, and \citetalias{green2019} are better than the remaining ones and, finally, \citetalias{g17} and 
GM20 are the best among the maps and models, respectively.

Table~\ref{corr} shows that, generally, the maps tend to give better predictions than the models do.
The reason is that the maps and models describe the dust medium by tables and formulas, respectively.
The former can take into account some short-term fluctuations of the medium, while the latter draw only a smooth pattern of the medium.
However, in contrast to maps, models allow us to derive the key parameters of the medium, as well as their relations and consequences.

\begin{table}
 \centering
\def\baselinestretch{1}\normalsize\normalsize
\caption[]{The correlation coefficients between the data and predictions from the models and maps.
}
\label{corr}
\begin{tabular}[c]{lcc}
\hline
\noalign{\smallskip}
 Model / map           & $mode(G_\mathrm{BP}-G_\mathrm{RP})$ & $mode(G_\mathrm{RP}-W3)$ \\
\hline
\noalign{\smallskip}
Simplest model                 & 0.52 & 0.54 \\
Model \citetalias{arenou}      & 0.76 & 0.75 \\ 
Model \citetalias{drimmel}     & 0.66 & 0.67 \\
Model \citetalias{av}          & 0.68 & 0.68 \\
Model GM20                     & 0.78 & 0.79 \\
Map \citetalias{g17}           & 0.87 & 0.86 \\
Map \citetalias{lallement2019} & 0.86 & 0.83 \\
Map \citetalias{green2019}     & 0.82 & 0.81 \\
\hline
\end{tabular}

\end{table}


\subsection{Comparison over the sky}

\begin{figure*}
\includegraphics{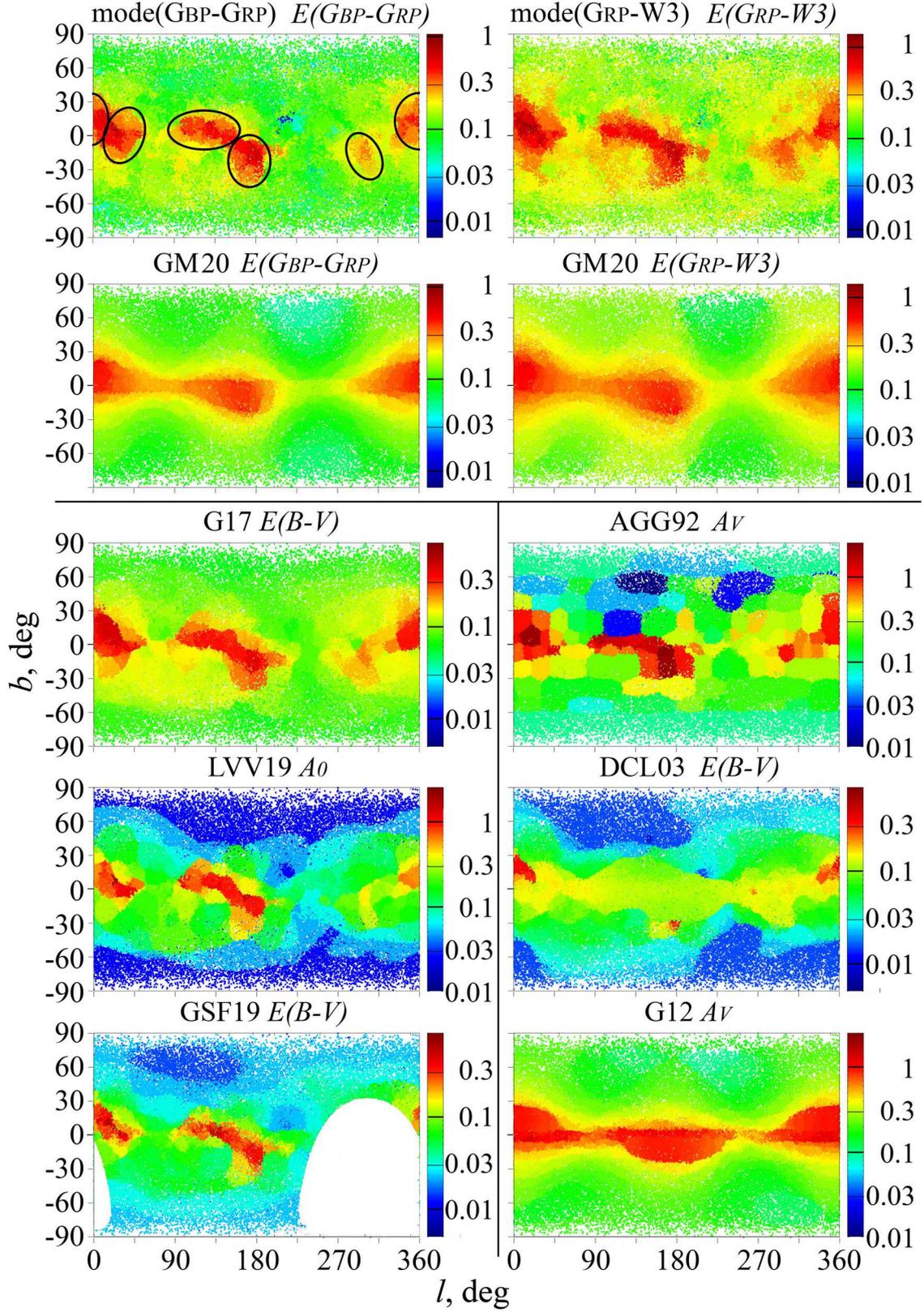}
\caption{The variations of the reddening/extinction from $mode(G_\mathrm{BP}-G_\mathrm{RP})$, $mode(G_\mathrm{RP}-W3)$ 
and various models and maps for 83\,317 cells over the sky in Galactic coordinates.
The five regions of a higher reddening are circled in the plot for the observed $mode(G_\mathrm{BP}-G_\mathrm{RP})$.
}
\label{lbmode}
\end{figure*}

The reddenings calculated from $mode(G_\mathrm{BP}-G_\mathrm{RP})$ and $mode(G_\mathrm{RP}-W3)$ for 83\,317 cells by use of 
relations~(\ref{bprpresult}) and (\ref{rpw3result}) are compared over the sky with the reddening or extinction estimates from 
various models and maps in Fig.~\ref{lbmode}.
All these reddening/extinction estimates are presented with the same scale based on the extinction law of \citetalias{wang2019}.

The upper row of the plots in Fig.~\ref{lbmode} shows the data, the second row shows the corresponding reddenings from GM20, 
then we show the predictions from three maps and three models in the left and right columns, respectively.

We arrange cells with similar $l$ and $b$ in Fig.~\ref{lbmode} to show the nearby ones as background, while more distant ones 
as foreground.
We select the size of the points in Fig.~\ref{lbmode} small enough in order to eliminate any obscuration of them at 
the middle and high latitudes.
Some obscuration exists only when the volume density of the cells is the highest, i.e. near the Galactic mid-plane, within 
$|b|<15\degr$.
However, even at $|b|<15\degr$ the uniform distribution of the stars and cells along $X$ and $Y$ leads to a dominance of 
the cells with a higher reddening, while the nearby cells with a lower reddening are rare and, hence, not important.

Fig.~\ref{lbmode} reveals the same five regions of a higher reddening/extinction as in Fig.~\ref{longi}, \ref{lgrw3}, 
and \ref{xy135}. 
They are circled in the plot for the observed $mode(G_\mathrm{BP}-G_\mathrm{RP})$.
These regions are drawn almost equally precise and detailed by the observed modes, \citetalias{g17}, \citetalias{lallement2019}, 
and \citetalias{green2019} (the region at $l\approx300\degr$ is not considered by \citetalias{green2019}),
but not by \citetalias{drimmel}.

The reason of this fail of \citetalias{drimmel} is evident from Fig.~\ref{lbmode}. It describes the dust distribution by 
a warped exponential disc with the addition of the local Orion arm features at $l\approx80\degr$ and $260\degr$.
However, a warp of the highly reddened regions of the disc, seen for \citetalias{drimmel} in Fig.~\ref{lbmode},
is actually the Belt layer tilted to the Galactic mid-plane.
The real warp of the equatorial dust layer, as shown in Sect.~\ref{longvar}, appears in the second and fourth Galactic quadrants 
at $R>400$~pc.
Moreover, the regions with a higher reddening at $l\approx80\degr$ and $260\degr$, drawn by \citetalias{drimmel} in 
Fig.~\ref{lbmode}, are absent in the data within 700~pc from the Sun.
These features were included into the \citetalias{drimmel} model following the features seen in some 2D dust emission maps 
at an unknown distance.
Probably they are related to some distant parts of the local arm, outside the space under consideration.

Fig.~\ref{lbmode} shows that, again, the maps give a more detailed and realistic picture than the models.

Note that \citetalias{arenou} shows some pixelization of the sky following the coordinate partitioning during the creation of 
this model.

Also, it is seen that GM20 draws a pattern closer to the data than the previous version \citetalias{av} does.

\section{Reddening across the whole dust layer}
\label{whole}

Different estimates of the reddening across the whole dust layer are discussed in \citetalias{gm2020}. 
Here we make some additional notes on this subject.

Fig.~\ref{lbmode} shows that the data, models and maps give different representation of the reddening/extinction at the middle and high latitudes.

In particular, in the upper row plots of Fig.~\ref{lbmode} the yellow colour for $E(G_\mathrm{RP}-W3)$ instead of the green 
colour for $E(G_\mathrm{BP}-G_\mathrm{RP})$ dominates at the middle and high latitudes.
This means that the observed $E(G_\mathrm{RP}-W3)$ exceeds $E(G_\mathrm{BP}-G_\mathrm{RP})$, 
when both the reddenings are derived from the data and normalized with the same extinction law of \citetalias{wang2019}.
This difference is not due to an offset. 
Hence, it cannot be changed by a change of dereddened colour~(\ref{bprpresult}) or (\ref{rpw3result}).
Instead, this difference is due to a wider range of the normalized $E(G_\mathrm{RP}-W3)$ w.r.t. the normalized 
$E(G_\mathrm{BP}-G_\mathrm{RP})$ at the middle and high latitudes. These ranges can be the same, if we use another extinction law.
Thus, this points to a slight deviation of the empirical extinction law from that of \citetalias{wang2019} at the middle and 
high latitudes\footnote{The extinction laws of \citetalias{davenport2014} for the high latitudes, \citetalias{wang2019}, 
\citetalias{schlafly2016}, and \citetalias{ccm89} with $R_\mathrm{V}=3.1$ give a similar slight deviation from the empirical law 
at the middle and high latitudes.}.
The same effect is seen for the reddening estimates from GM20 (the second row of the plots in Fig.~\ref{lbmode}).
The difference between the GM20 plots in Fig.~\ref{lbmode} corresponds to the difference between $Z_A$ for 
$G_\mathrm{BP}-G_\mathrm{RP}$ and $G_\mathrm{RP}-W3$ in Table~\ref{solution}.
All these findings may suggest a slightly different vertical spatial distribution of the fine and coarse dust grains.

The colour of the middle and high latitudes in Fig.~\ref{lbmode} allows us to rank the maps and models in ascending order of 
their estimates of low extinctions/reddenings as:
\citetalias{lallement2019}, \citetalias{drimmel}, \citetalias{green2019}, and \citetalias{arenou} -- show an apparent 
underestimation of the low extinctions/reddenings w.r.t. the observed ones in the upper row plots -- versus \citetalias{av}, 
GM20, and \citetalias{g17}, which agree with each other and with the observed values.
This diversity of the estimates of low reddening, including the reddening across the whole dust layer, is noted in \citetalias{gm2020}.

\begin{figure*}
\includegraphics{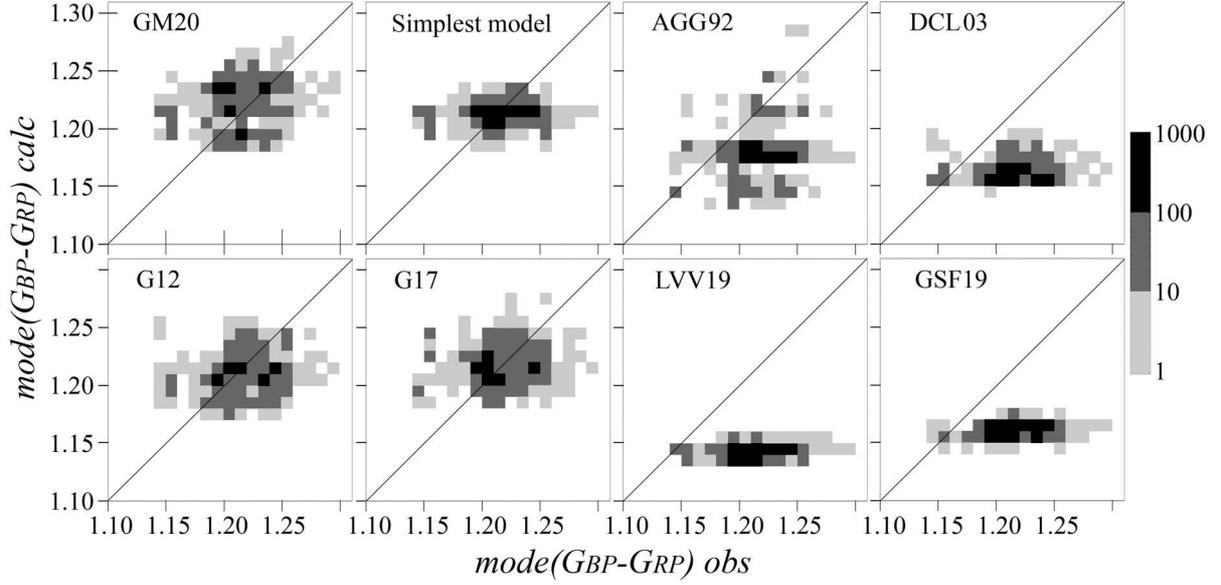}
\caption{The same as Fig.~\ref{gbgrmode} but for 2788 cells within $(X^2+Y^2)^{0.5}<120$~pc.
The number of the stars in each bin of $0.01 \times 0.01$ mag is shown by the colour scale on the right.
}
\label{gbgrmodexy120pc}
\end{figure*}

We compare in Fig.~\ref{gbgrmodexy120pc} several 3D reddening models and maps in their ability to predict the observed 
$mode(G_\mathrm{BP}-G_\mathrm{RP})$ for the 2788 cells within $(X^2+Y^2)^{0.5}<120$~pc, i.e. within a vertical 
cylinder across the whole dust layer.
This cylinder is quite narrow to contain as unreddened nearby stars, as the stars behind the whole dust layer.
A similar figure can be shown for $mode(G_\mathrm{RP}-W3)$.

The vertical offsets of \citetalias{arenou}, \citetalias{drimmel}, \citetalias{lallement2019}, and \citetalias{green2019} from 
the bisector line in Fig.~\ref{gbgrmodexy120pc} can be corrected by accepting an intrinsic colour higher than (\ref{bprpresult}).
However, as noted in Sect.~\ref{compar}, this cannot correct any prediction of the colour range of the observed 
$mode(G_\mathrm{BP}-G_\mathrm{RP})$, i.e. the accordance of the abscissa and ordinate ranges in Fig.~\ref{gbgrmodexy120pc}.
The abscissa range reflects an observed diversity of the colours due to the reddening across the whole dust layer.
Hence, the best predictions should cover the same colour range along the ordinate.
However, Fig.~\ref{gbgrmodexy120pc} shows that the simplest model, \citetalias{drimmel}, \citetalias{lallement2019}, and 
\citetalias{green2019} cover too narrow colour ranges (i.e. their ranges along the ordinate).
The only way to make their colour ranges wider is a higher reddening.
Thus, Fig.~\ref{gbgrmodexy120pc} confirms that the simplest model, \citetalias{drimmel}, \citetalias{lallement2019}, and 
\citetalias{green2019} underestimate the reddening across the whole dust layer.
The same is true for a similar figure of $mode(G_\mathrm{RP}-W3)$.

Some models and maps show rather tight correlations of their reddenings between each other within the whole space cylinder 
under consideration
[the correlation coefficients for $mode(G_\mathrm{BP}-G_\mathrm{RP})$ and $mode(G_\mathrm{RP}-W3)$ are averaged]:
\begin{itemize}
\item 0.91 for GM20 versus \citetalias{av}, 
\item 0.86 for \citetalias{g17} versus \citetalias{lallement2019},
\item 0.84 for \citetalias{lallement2019} versus \citetalias{green2019},
\item 0.83 for GM20 versus \citetalias{g17},
\item 0.82 for \citetalias{g17} versus \citetalias{green2019}.
\end{itemize}
Similar tight correlations have been found by \citet{polarization} between \citetalias{g17}, \citetalias{av}, \citetalias{arenou}, 
on the one hand, and a previous (albeit rather similar) version \citetalias{lallement2014} of \citetalias{lallement2019}, on 
the other hand.
This means that, at least, the independent reddening estimates from \citetalias{g17}, \citetalias{lallement2019}, and 
\citetalias{green2019} can be basically described as some linear combinations of each other (this is also evident from 
Fig.~\ref{gbgrmode}). 
In particular, \citet{polarization} have shown in their figure 11 and their equation (4) that \citetalias{g17} differs from 
\citetalias{lallement2014} mostly by the constant offset of $\Delta E(B-V)=0.06$ mag.
This offset can be interpreted as the difference between the estimates of the reddening across the whole dust half-layer
from \citetalias{g17} and \citetalias{lallement2014}.
Indeed, a similar offset of $\Delta E(G_\mathrm{BP}-G_\mathrm{RP})=0.07$ mag [$\Delta E(B-V)=0.05$ mag with the extinction 
law of \citetalias{wang2019}]
between the predictions of \citetalias{g17} and \citetalias{lallement2019} is seen in Fig.~\ref{gbgrmodexy120pc}.
Such a constant offset means that the difference between the maps/models is more noticeable for low reddenings, 
i.e. near the Sun and at high latitudes.
This offset also means that, if \citetalias{g17} overestimates $E(G_\mathrm{BP}-G_\mathrm{RP})$ by 0.01 mag, then 
\citetalias{lallement2019} underestimates it by 0.06 mag [by $\Delta E(B-V)=0.045$ mag with the extinction law of 
\citetalias{wang2019}].

Table~\ref{summ} summarizes the comparable estimates of the reddening across the whole dust half-layer below or above the Sun 
from \citetalias{gm2020} and the current paper.  They agree with each other. 
The estimates from \citetalias{gm2020} and GM20 converge to the average value $E(B-V)=0.059\pm0.003$ mag by use of the extinction 
law of \citetalias{wang2019}.
Note that these new estimates are lower than that from the previous version of the model, \citetalias{av}, also presented 
in Table~\ref{summ}.
However, the diversity of the estimates from different maps and models, summarized in table~6 of \citetalias{gm2020}, 
shows that the estimation of the reddening across the whole dust half-layer is still an issue.

\begin{table*}
 \centering
\def\baselinestretch{1}\normalsize\normalsize
\caption[]{Some estimates of the extinction or reddening across the whole dust half-layer below or above the Sun.
}
\label{summ}
\begin{tabular}[c]{lcc}
\hline
\noalign{\smallskip}
 Source           & Original estimate & $E(B-V)$ by \citetalias{wang2019} \\
\hline
\noalign{\smallskip}
\citetalias{gm2020}, narrow cylinder      & $E(G_\mathrm{BP}-G_\mathrm{RP})=0.080$ & 0.061 \\
This study, simplest model       & $E(G_\mathrm{BP}-G_\mathrm{RP})=0.068$ & 0.051 \\
This study, model GM20           & $E(G_\mathrm{BP}-G_\mathrm{RP})=0.073$ & 0.055 \\
\hline
\noalign{\smallskip}
\citetalias{gm2020}, narrow cylinder      & $E(G_\mathrm{RP}-W3)=0.100$   & 0.058 \\
This study, simplest model       & $E(G_\mathrm{RP}-W3)=0.090$            & 0.052 \\
This study, model GM20           & $E(G_\mathrm{RP}-W3)=0.105$            & 0.061 \\
\hline
\noalign{\smallskip}
Model \citetalias{av}            & $A_\mathrm{V}=0.214$                   & 0.068 \\
\hline
\end{tabular}

\end{table*}


\section{Conclusions}
\label{conclusions}

In this study we exploited the complete sample of 101\,810 {\it Gaia} DR2 giants.
It has been selected in \citetalias{gm2020} within the clump domain of the Hertzsprung--Russell diagram and in the space cylinder 
with the radius 700~pc around the Sun, up to $|Z|<1800$~pc along the $Z$ Galactic coordinate.
Owing to the accurate parallaxes and photometry in the $G_\mathrm{BP}$ and $G_\mathrm{RP}$ bands from {\it Gaia} DR2, in 
combination with the {\it WISE} photometry in the $W3$ band, this sample is the first one, which is complete for the space across 
the whole Galactic dust layer near the Sun.

In the current paper, we presented GM20, the revised version of the 3D dust distribution model introduced by Gontcharov in 2009.
We used GM20 to explain the spatial variations of the colour modes $mode(G_\mathrm{BP}-G_\mathrm{RP})$ and $mode(G_\mathrm{RP}-W3)$.
This allowed us to derive simultaneously the key properties of the Galactic dust layer, the intrinsic colours and absolute 
magnitudes of the giant clump, and vertical gradients of the intrinsic colours with $|Z|$.

We revealed five regions with high reddening $mode(G_\mathrm{BP}-G_\mathrm{RP})>1.35$ and $mode(G_\mathrm{RP}-W3)>1.75$.
The regions at $l\approx30\degr$, $120\degr$, and $300\degr$ contain the Aquils South, Polaris Flare, and Chamaeleon cloud complexes, 
respectively, and belong to the equatorial dust layer.
Other ones at $l\approx5\degr$ and $175\degr$ contain Aquila rift, Ophiuchus, Lupus and Taurus -- Perseus -- Auriga, Orion cloud 
complexes, respectively, and belong to the dust layer of the Gould Belt inclined to the Galactic mid-plane at about $17\degr$.

Taking into account this analysis, we proposed for GM20 the 3D dust distribution in two intersecting dust layers, 
with exponential vertical and sinusoidal longitudinal variations of the dust spatial density in each layer.
Twenty free parameters of our GM20 model were calculated for the least root-mean-square deviation of this model from the data.
$mode(G_\mathrm{BP}-G_\mathrm{RP})$ and $mode(G_\mathrm{RP}-W3)$ gave similar solutions, but slightly different equatorial layer 
scale heights of $150\pm15$ and $180\pm15$~pc, respectively.
The current solution for GM20 presents the Belt layer as an ellipse, oriented nearly between the centre and anticentre of the Galaxy, 
with a maximum reddening in roughly these directions, and with a semi-major and -minor axes of 600 and 146~pc, respectively.
This agrees well with the known structure of the Belt.
Note that the Belt dust container produces a considerable extinction at the middle and high latitudes.

GM20 estimates for nearby clump giants $(G_\mathrm{BP}-G_\mathrm{RP})_0=(1.14\pm0.01)-(0.022\pm0.010)\,|Z|$ and 
$(G_\mathrm{RP}-W3)_0=(1.44\pm0.01)-(0.015\pm0.010)\,|Z|$, where $Z$ is expressed in kpc, 
agree with the estimates from Paper I, obtained in a different approach.

Also, GM20 estimates agree with the PARSEC, MIST and IAC-BaSTI theoretical isochrones for the clump giants, with their expected 
median age of 3 Gyr and median metallicity of $\mathrm{[Fe/H]}=-0.11$.
The only exception is the MIST isochrone based on the MIST predictions for $W3$.
It deviates, probably, due to an issue in its colour--$T_\mathrm{eff}$ relation and/or bolometric correction used.

We compared the observed $mode(G_\mathrm{BP}-G_\mathrm{RP})$ and $mode(G_\mathrm{RP}-W3)$ with the predictions from GM20 and 
several 3D reddening models and maps. GM20 and \citetalias{g17} appeared to be the best 3D reddening model and map, respectively.

All our results from \citetalias{gm2020} and from this study agree with the reddening across the whole dust half-layer 
below or above the Sun $E(B-V)=0.059\pm0.003$ mag.

\citetalias{drimmel}, \citetalias{lallement2019}, and \citetalias{green2019} need much higher giant clump intrinsic colours
to reduce their residuals with the data.
Such colours cannot explain as much data with the lower observed dereddened colours.
Thus, we are inclined to conclude that \citetalias{drimmel}, \citetalias{lallement2019}, and \citetalias{green2019}
systematically underestimate low reddenings.
This is confirmed by the too narrow range of their predictions for the colours inside the vertical space cylinder across the whole 
dust layer.

However, \citetalias{g17}, \citetalias{lallement2019}, and \citetalias{green2019} show such high linear correlation 
coefficients in the whole space under consideration that their reddening predictions differ almost exclusively by constant offsets.
These offsets might arise from their different estimates for the low reddenings near the Sun and across the whole dust half-layer.
Thus, a correct estimation of such low reddenings, as well as the creation of the samples of unreddened stars, 
seem to be the key points for the modern investigations of extinction and reddening in the Galaxy.

\section*{Acknowledgements}

We thank the reviewer, Dr. Xiaodian Chen for useful comments.
We thank Jeremy Mutter for assistance with English syntax.
We thank
Anton Dryanichkin,
Evgeny Evseev,
Marina Garanina,
Evgeny Gordeev,
Pavel Popov, and
Natalia Svetlova for their help in calculations.

We acknowledge financial support from the Russian Science Foundation (grant no. 20--72--10052).

The research described in this paper makes use of Filtergraph \citep{filtergraph}, 
an online data visualization tool developed at Vanderbilt University through the 
Vanderbilt Initiative in Data-intensive Astrophysics (VIDA) and the Frist Center for Autism and Innovation (FCAI, \url{https://filtergraph.com}).
The resources of the Centre de Donn\'ees astronomiques de Strasbourg, Strasbourg, France
(\url{http://cds.u-strasbg.fr}), including the SIMBAD data base and the X-Match service, were widely used in this study.
This work has made use of data from the European Space Agency (ESA) mission {\it Gaia}
(\url{https://www.cosmos.esa.int/gaia}), processed by the {\it Gaia} Data Processing and Analysis Consortium
(DPAC, \url{https://www.cosmos.esa.int/web/gaia/dpac/consortium}).
This publication makes use of data products from the {\it Wide-field Infrared Survey Explorer}, which is a joint project
of the University of California, Los Angeles, and the Jet Propulsion Laboratory/California Institute of Technology.

\section*{Data availability}

The data underlying this article are available in the article and in its online supplementary material.

\bsp	
\label{lastpage}
\end{document}